\documentclass{article}

\usepackage{arxiv}

\usepackage[utf8]{inputenc} % allow utf-8 input
\usepackage[T1]{fontenc}    % use 8-bit T1 fonts
\usepackage{hyperref}       % hyperlinks
\usepackage{url}            % simple URL typesetting
\usepackage{booktabs}       % professional-quality tables
\usepackage{amsfonts}       % blackboard math symbols
\usepackage{nicefrac}       % compact symbols for 1/2, etc.
\usepackage{microtype}      % microtypography
\usepackage{lipsum}		% Can be removed after putting your text content
\usepackage{graphicx}
\usepackage{natbib}
\usepackage{doi}
\usepackage{multirow}

\title{Simulation of emergency evacuation of passengers with and without disability at different types of metro stations}

\author{ \href{https://orcid.org/0000-0002-5460-1034}{\includegraphics[scale=0.06]{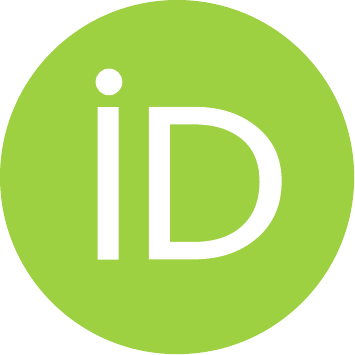}\hspace{1mm}Tarapada~Mandal} \\
	Department of Civil Engineering\\
	Indian Institute of Technology Delhi\\
	Hauz Khas, Delhi 110016 \\
	\texttt{tarapada@civil.iitd.ac.in} \\
	\And
	\href{https://orcid.org/0000-0002-7229-519X}{\includegraphics[scale=0.06]{orcid.pdf}\hspace{1mm}K. Ramachandra~Rao} \\
	Department of Civil Engineering\\
	Indian Institute of Technology Delhi\\
	Hauz Khas, Delhi 110016  \\
	\texttt{rrkalaga@civil.iitd.ac.in} \\
        \And
	\href{https://orcid.org/0000-0002-9302-1243}{\includegraphics[scale=0.06]{orcid.pdf}\hspace{1mm}Geetam~Tiwari} \\
	Department of Civil Engineering\\
	Indian Institute of Technology Delhi\\
	Hauz Khas, Delhi 110016\\
	\texttt{geetamt@gmail.com} \\
}

\renewcommand{\shorttitle}{\textit{arXiv} Template}

\hypersetup{
pdftitle={Simulation of emergency evacuation of passengers with and without disability at different types of metro stations},
pdfsubject={q-bio.NC, q-bio.QM},
pdfauthor={Tarapada~Mandal, K. Ramachandra~Rao, Geetam~Tiwari},
pdfkeywords={Assised evacuation, Simulation, Metro station},
}

\begin{document}
\maketitle

\begin{abstract}
	\ Metro systems are part of major transportation systems for big cities. Evacuation is a key challenge for metro systems in case of fire or terrorist attacks. In case of evacuation, wheelchair-assisted evacuees might take a longer time. In order to understand the effect of assisted and non-assisted evacuees on evacuation, a simulation is conducted in this study. Platform evacuation and train evacuation simulation are done. A train load survey is conducted to understand the number of evacuees inside a train. Two different layouts of stations are considered, underground island-type platforms and elevated platforms. Simulation of wheelchair-assisted and non-assisted evacuees is done. The design of the experiment is used to create full factorial and fractional factorial designs to take input of different factors in simulation. The main factors affecting the total evacuation time are calculated using the design of the experiment. It is found that wheelchair-assisted evacuees take longer time than non-assisted evacuees on both underground and elevated station platforms. It is also found that efforts are needed to increase the speed of the non-assisted evacuees also.
\end{abstract}

% keywords can be removed

\section{Introduction}
\ There are many threats to urban life. Transportation systems (underground or elevated metro systems) are also under constant threat of terrorist attacks and fire incidents. In order to provide safety to the commuters, we need a safe evacuation plan. Evacuation plans often describe the wayfinding process in a metro station. In other words, they describe the paths those need to be followed to reach the exits (those leads to safety) in case of an emergency evacuation. To get commuters accustomed to the evacuation plans the safety managers or engineers need to conduct evacuation drills which becomes impossible as the number of commuters is very high. There are ethical issues with conducting drills as people may get injured. The cost of conducting such drills is also high. So, the safety managers perform experiments using computer simulation to understand or assess different situations during real evacuations. Emergency evacuation simulation models are generally description models that can give a description of what happens in an emergency situation. Environmental effects (dimension of the infrastructure, position of fire, spread of fire and smoke) and human behaviour play key roles in evacuation dynamics. So, any emergency evacuation simulation must incorporate as many factors as possible from environmental components as well as human behaviour. Although different behaviours (attitudinal) during metro station evacuation have been suggested by one recent study \citep{Shiwakoti2017}, only a few quantifiable behavioural parameters are taken for simulation purposes in general. Different parameters for human behaviours could be varying levels of speed and pre-movement times due to different levels of perceived threats (as the cognitive power of individuals is different). Many simulation studies (as discussed in section 2) for metro stations have used different parameters (speed, pre-movement time, number of evacuees etc.) for simulation in different types of models (social force, cellular automata, agent-based).  In this current study, the design of experiments is used to understand the effects of different parameters (and their 2nd and 3rd order interactions) on evacuation time. Delhi Metro like other world-class public metro systems is disabled-friendly. So, it is expected that a considerable number of disabled commuters use Delhi Metro.  Simulations for assisted evacuation in different layouts (underground and elevated) of Delhi Metro stations are conducted. To our knowledge, assisted evacuation for metro stations using the design of experiments for simulation of evacuation is not done in any studies. Section 2 provides a detailed literature review on evacuation simulation studies on metro stations. Section 3 provides the layouts of the metro stations as well as the pathfinder model. Section 4 illustrates the choice of different simulation parameters. Section 5 discusses about the design of experiments for various simulation cases with results. Statistical analysis is also done for various experiments. The last section (section 6) is for the conclusion.

\section{Literature Review}

\subsection{Non-assisted evacuation simulation models}
\ Although there are review papers for simulation models \citep{Kuligowski2005} and optimization models \citep{Vermuyten2016} discussing pedestrian emergency evacuation in buildings, not much importance is given to metro station evacuation simulations. For normal (non-assisted) evacuation simulation some interesting studies are found in the literature. These are discussed in this section. Some of the studies \citep{Chen2017,MengnJia2017,wang2013,yin,Li2014,Zhang2016} have used pedestrian module of Any Logic software which is based on social force model of pedestrian dynamics. Some studies \citep{Jiang2010,Shi2012} have used building EXODUS \citep{owen}. Some studies \citep{Hong2016,Zhong2008} have developed their own model (network-grid based) for simulation. None of these studies could incorporate assisted evacuation in simulation. Few have considered fire scenarios (location of fire, range of fire). A detailed analysis of parameters taken in these models has been illustrated in Table 1.
%\begin{equation}
%	\xi _{ij}(t)=P(x_{t}=i,x_{t+1}=j|y,v,w;\theta)= {\frac {\alpha _{i}(t)a^{w_t}_{ij}\beta _{j}(t+1)b^{v_{t+1}}_{j}(y_{t+1})}{\sum _{i=1}^{N} \sum _{j=1}^{N} \alpha _{i}(t)a^{w_t}_{ij}\beta _{j}(t+1)b^{v_{t+1}}_{j}(y_{t+1})}}
%\end{equation}

\subsection{Assisted Evacuation simulation models}
\ Metro organizations worldwide make everything possible to make their facility points (train, station, and platform) more accessible for disabled commuters. For emergency evacuation from a metro system, it is important to consider the safe evacuation of disabled commuters. There is a huge lack of real empirical data for evacuation simulation particularly assisted evacuation simulation. To our knowledge, no studies have been done on assisted evacuation in metro stations. Table 2 provides parameter information about assisted evacuation in other infrastructures (hospital, institutional buildings, and residential buildings) found in the literature. 
%%table1
\begin{table}
        \caption{Parameters taken for evacuation in Metro Station}
        \centering
        \small
        \resizebox{\columnwidth}{!}{\begin{tabular}{|c|c|c|c|c|c|c|}
            \hline
                Authors & Speed in platform (m/s) & Speed in stairs (m/s) & Number of evacuees & Pre-evacuation time (sec) & Assisted evacuation? & Fire Scenarios considered? \\
            \hline
                \citep{MengnJia2017} & 1.61 & 0.65 & 1100+370 & 60 & No & No\\
            \hline
                \citep{Zhong2008} & 1.2-1.5 & 0.6-0.67 & 1300 & 60 & No & No \\
            \hline
                \citep{Shi2012} & 0.9-1.5 & 0.4-0.7 & 1336+554 & 60 & No & No \\
            \hline
                \citep{Chen2017} &  1.10-2.05 & 0.66-1.23 & 3360 & 60 & No & No \\
            \hline
                \citep{Zhang2016} &  - & - & 550-750 & 60 & No & No \\
            \hline
                \citep{Jiang2010} &  - & 0.67-0.79 & 1168-5172 & 0-60, 60-90, 60-120 & No & No \\
            \hline
                \citep{Li2014} &  - & - & 356-2756 & 60 & No & No \\
             \hline
                \citep{wang2013} &  0.98-2.15 & 0.98-2.15 & 800 & 60 & No & No \\
            \hline
                \citep{Hong2016} &  1.57,1.17 & 1.57,1.17 & - & 0 & No & No \\
            \hline
                \citep{yin} &  - & - & - & - & No & No \\
            \hline
        \end{tabular}}
    \end{table}

%%table2
\begin{table}
        \caption{Parameters taken for evacuation in Metro Station}
        \centering
        \small
        \resizebox{\columnwidth}{!}{\begin{tabular}{|c|c|c|c|c|c|c|c|c|c|}
            \hline
                Authors & Speed of wheelchair on level surface (m/s) & Speed on level surface of others (m/s) & Speed on stairs of others (m/s) & Number of evacuees & Number of evacuees with disability & Number of people assisting & Pre-evacuation time (Sec) /preparation time & Built environment case & Smoke effect considered? \\
            \hline
                \citep{christensen2008} & 0.89 & 1.25 & 0.70 & 71 & 6 & 0 & - & Research building & No\\
            \hline
                \citep{Geoerg2016} & - & (0.30-0.60), (0.8-1.2), (0.33-0.8) & - & 100 & (13+21) & 0 & (120, 60, 180) & Room evacuation & No\\
            \hline
                \citep{Levin1989} & - & 1.69, 0.65, 0.845, 0.78, 0.52 & - & 2 & 1 & 1 & 90 & House evacuation & Yes\\
            \hline
                \citep{Rahouti2016} & 1.46 & 1 & - & 28 & 14 & 6 & 50.8; 32.7 & Hospital evacuation & No\\
            \hline
                \citep{Johnson2005} & - & 1.12, 1.68, 0.52 & - & 30 & 15 & 10 & - & Hospital evacuation & No\\
            \hline
                \citep{yokouchi2017} & 2.46 & 1.20 & - & 60 & 6, 24, 48; 34  & 3 to 60 & 26.1; 14.1 & Hospital evacuation & No\\
            \hline
                \citep{Konstantara2016} & - & 1.34 & - & 110 & 30 & 0, 8, 11, 15, 23, 30 & - & Old age home floor evacuation & No\\
            \hline
        \end{tabular}}
    \end{table}

\section{Station Layouts and PATHFINDER Model}
\subsection{Station layout}
\ Delhi Metro is the largest metro in India. It has currently eight different lines (Blue, Yellow, Red, Orange, Violet, Green, Pink, and Magenta) running across Delhi and NCT-Delhi (national capital territory). It is having underground and elevated platforms for different lines. In our simulation, we have chosen one underground and one elevated platform layout of a typical yellow line station. More information about the fire locations for the underground and elevated stations have been discussed in detail in section 4 and section 5. Figure 1 and Figure 2 are the typical sketches for an underground and an elevated Delhi metro station. Delhi metro has underground and elevated stations at different depths and heights. In our simulation, we have considered some typical layouts of the Delhi Metro. We have used 3D geometry (cross-sectional) of typical Delhi Metro stations and imported it into PATHFINDER for evacuation simulation. Table 3 provides typical dimensions of an underground and an elevated station (considered for simulation) of the Delhi Metro.

\begin{figure}
\centering
\includegraphics{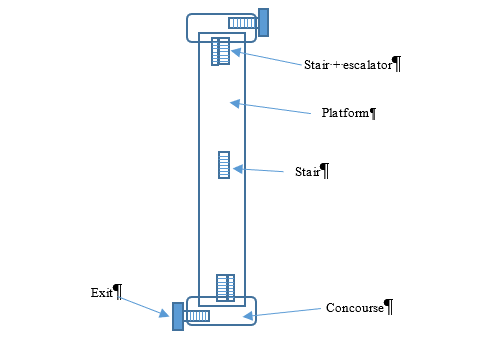}
\caption{Typical sketch of an underground island type station of Delhi Metro}
\end{figure}

\begin{figure}
\centering
\includegraphics{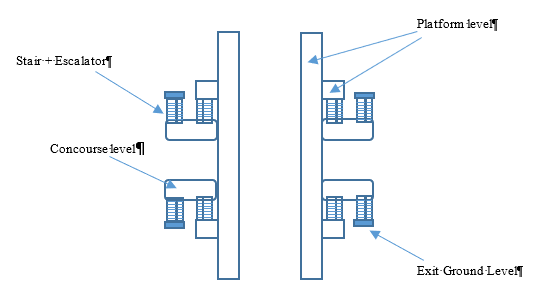}
\caption{Typical sketch of an elevated station of Delhi Metro }
\end{figure}

%%table3
\begin{table}
        \caption{Characteristics of the two typical Delhi Metro Station layouts}
        \centering
        \small
        {\begin{tabular}{|c|c|c|}
            \hline
                Characteristics & Underground (in meters) & Elevated (in meters)\\
            \hline
                Length of Platform & 185 & 185 \\
            \hline
                Width of platform & 8 & 4 on each side \\
            \hline
                Height/Depth of platform & -12 & 9 \\
            \hline
               Width of stair & 3 & 3 \\
            \hline
               Width of Escalator & 1.1 & 1.1 \\
            \hline
        \end{tabular}}
    \end{table}
    
\subsection{Pathfinder Model}
\ PATHFINDER is an agent-based specialized evacuation simulation software developed by Thunderhead Engineering. Most of the parameters of the pathfinder are calibrated according to the guidelines of the SFPE (Standard for Fire Protection Engineers). To our knowledge, different types of metro station layouts have not been simulated for understanding evacuation in any studies till now using PATHFINDER. Metro stations have complex geometry and the critical points are the intersections of horizontal and vertical infrastructures (platform and stairs). So, the speed density profile plays a significant role in evacuation on those infrastructures.  Choosing the correct speed-density profile is important to understanding the evacuation process. We do not have a speed-density relationship in emergency evacuation situations due to a lack of data (as it is impossible to collect real emergency evacuation data). Taking a speed-density profile from normal crowd flow movement to simulate emergency evacuation (as done in most of the studies) can be wrong. So, we have used SFPE profile of the speed-density relationship in our study. The desired speed values (as inputs as discussed in section 4) are changed according to the density of the surrounding population on any type of terrain (platform, stairs, etc.) as mentioned in the SFPE handbook (2003, 2016), in emergency movement chapters. Though there are two versions of SFPE mentioned here, the main contribution was made by Nelson and Mowrer \citep{Nelson2003} and followed in the latest version by Gwynne and Rosenbaum \citep{Gwynne2016}. So, using SFPE speed density profile (as in our model) is more justified in fire emergency movement cases. As PATHFINDER provides the platform to use different speed density profiles as mentioned in the SFPE guidelines, it can be one of the best tools for simulating emergency evacuation.

\section{Choice of Simulation Parameters}
\ Evacuation simulation experiment is carried out in some of the typical layouts of both underground and elevated stations of yellow line of Delhi Metro. Delhi metro yellow line is the 2nd longest line in Delhi Metro. It is having 17 elevated and 20 underground stations. All the elevated stations have side platforms. Some of the underground stations (five) have side platforms while some of them (fifteen) have island platform layouts. In this study we have not only considered different types of station layouts but also different types of fire cases. We have chosen two different station layouts for this study.
\begin{itemize}
	\item Underground station (island platform) (refer Figure 1)
	\item Elevated station (side platform) (refer Figure 2)
\end{itemize}
The main two fire cases for the underground station are
\begin{itemize}
	\item Fire on the middle of the platform
	\item Fire on the middle of the train
\end{itemize}
\ The main fire location for the elevated station is fire near to one of the stair exits on platform. The reason behind choosing fire location in the middle of the platform and middle of the train for underground station is that middle positions are more critical positions and they create more vulnerable situations for the evacuees. The fire position for the elevated station is chosen at one of the exits because for elevated stations there are two exits and blocking one exit could lead to a critical situation.
We have also considered simulation for the disabled people as public transit stations are designed as disabled-friendly. To our knowledge, our study is the first one to consider disabled people in transit station evacuation simulation. So we have two base cases considering different types of people in our simulations:
\begin{itemize}
	\item Simulation with people with no need of assistance
	\item Simulation with people with need for assistance and no need for assistance
\end{itemize}
\subsection{Simulation parameters for non-assisted evacuation case}
\ From the literature study in section 2 we found the influencing factors in evacuation are:
\begin{itemize}
	\item Number of evacuees
	\item Speed of evacuee
	\item Width of staircase 
        \item Pre-evacuation time
\end{itemize}
\subsection{Simulation parameters for assisted evacuation simulation case}
\ From the literature we found no study that has been done on assisted evacuation simulation of a transit station. But based on the studies done on assisted evacuation simulation on other built-in spaces (hospitals, residential buildings etc.), we have chosen the parameters for assisted evacuation simulation in our study. The following are the different parameters considered for assisted evacuation simulation experiment in different case scenarios in two different metro station layouts.
\begin{itemize}
	\item Number of evacuees
	\item Speed of evacuee
	\item Number of disabled people
        \item Speed of disabled people
        \item Pre-evacuation time
        \item Width of stair
\end{itemize}
\subsection{Parameter ranges for two types of simulation cases}
\ In this study the parameters are kept as realistic as possible. We have taken two different levels of each parameter. We also have used design of experiment method for generating inputs for simulation of different cases. To our knowledge this is the first study to use design of experiment (a method used in Industrial Engineering to study effect of different factors on process yield) in simulation study. The method is discussed in details in section 5. To our knowledge this is the first study to conduct different ancillary studies (surveys) to get the simulation parameters as realistic as possible. 
\subsubsection{Passenger load profile survey}
\ To make the simulation realistic, supplementary surveys are conducted in Delhi Metro to assess the number of people on the platform as well as inside trains for different metro stations. During peak hours passenger load survey is conducted inside trains of yellow line. Passengers on the platforms are obtained from the ridership survey (how many people are entering a station). Passengers inside a train is crucial for train evacuation and this number is obtained from the train load survey. Survey started at one end of the yellow line with the count of people inside the train and at each station boarding and alighting passengers are counted to find the loads at each station. Figure 3 provides a load profile of number of people inside a train for each station.
As we are not simulating any transfer stations which comes under the red section due to huge passenger loads (we do not have geometry details of the transfer stations yet) and we are considering for underground stations with island type platforms, the train passenger loads for these types of stations varies between 1400 and 1650. When we look for the elevated stations, this numbers vary between 1000 and 1500. So, we have taken these values for simulation. It is also found from the ridership details that number of people on a platform in between two consecutive trains is varying between 250 and 450 for underground island type platforms. The value is between 150 and 400 for elevated stations. In the next section (design of experiments and results) we have showed the input parameters for different experiments based on the surveys done.
\begin{figure}
\centering
\includegraphics[width=\textwidth]{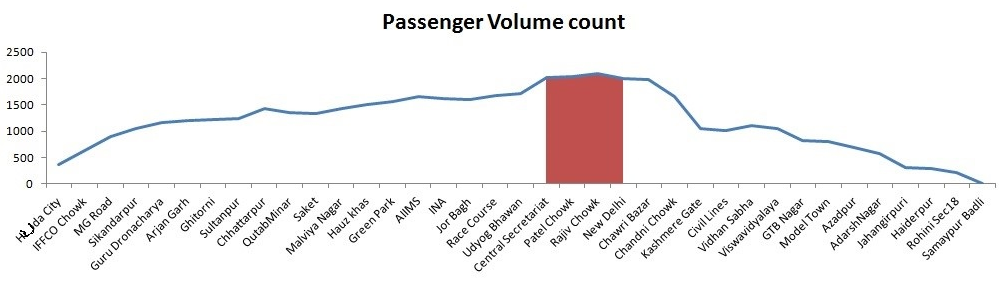}
\caption{Passenger load profile inside train at each station of yellow line}
\end{figure}
\subsubsection{Speed choice for people with no disabilities}
\ Evacuee’s speed is one of the major factors in evacuation dynamics. The station architecture as well as the evacuation plans should be drafted in such a way so that the commuters can always attain a maximum unimpeded free-flow speed during evacuation. Free flow speed can be different for different evacuees based on their perceived level of panic. The threshold value of evacuee speed above which irregular patterns (arching at bottlenecks) can be found in the crowd stream is 1.5 m/s as mentioned in the seminal work by Helbing \citep{Helbing2000}. It is widely accepted and irregular patterns are common in escaping crowd. Another widely accepted study to assess speed of escaping people is the study of Predtechenskii and Milinskii \citep{PredTechenskii1978} where the authors suggested that the speed at normal situation can be multiplied by different factors (for corridors, upstairs and downstairs) to obtain speeds for emergency evacuation. 
According to the station planning standards of Delhi Metro, the average walking speed of passengers on different infrastructures is 1.0 m/s. So, for the lower range (of the two-level design of experiment) we have kept the speed value 1.0-1.5 m/s (keeping in mind the threshold value of emergency evacuation in Helbing’s paper) with a uniform distribution among the evacuee population.  Helbing’s simulation (social force model) study suggested that evacuation speed of panicked evacuees can reach more than 5 m/s. In their simulation study they showed that injuries occur when the free flow speed exceeds 5 m/s. An agent-based simulation study \citep{Ha2012} also finds irregular variations in evacuation time when the desired speed exceeds 5 m/s. They also found that evacuation time is minimised when the desired speed is chosen as 2 m/s. Another study \citep{wang2013} has chosen varying range of speed value (1.1-2.9) for simulation of different fire evacuation cases in a metro station in China. One empirical study \citep{Kholshevnikov2008} in Russia suggests that under stress condition (due to panic) the free-flow speed is 1.5-2.0 m/s on levelled corridors as well as stairs. In our study higher value of speed is taken as 1.5-2.15 m/s (uniform distribution). 
\subsubsection{Number of disabled evacuees}
\ We have used the data from Census 2011 and NSS (national sample survey) for choosing the disabled evacuee number as a simulation parameter in assisted evacuation cases. According to census 2011 and NSS, 1.4 percent of the total population of NCT (National Capital territory) Delhi are disabled. It is not possible to get the exact figures of how many disabled people use the Delhi-Metro from Census 2011 and NSS data. We could not get any data from Delhi Metro. So, we have taken number of disabled evacuees as 1.4 percent of the total evacuee numbers or less for different cases of assisted evacuation simulation. We also found from the NSS data that 28 percent of the disabled people are working. So, there is a chance that at any time they might use public transport like Delhi Metro. We have used two different levels for this parameter (lower value: keeping in mind the current situation as very few fully disabled people use Delhi Metro and higher value: keeping in mind that Delhi metro is disabled friendly and more disabled people are encouraged to use the Delhi Metro). According to our knowledge, only PATHFINDER and buildingEXODUS are the two-simulation software which can simulate assisted evacuation realistically.  So, in our PATHFINDER model we have done realistic assisted evacuation of Delhi Metro. However, we are only considering people in a wheelchair. People with assistive walker are not considered in the current simulations. The different values of this parameter are shown in section 5 (design of experiment).
\subsubsection{Speed of disabled evacuees}
\ Realistic choice of this parameter is tricky. To our knowledge, no studies (empirical, experiment or simulation) have been done on assisted evacuation in Metro Stations. We have found some studies on different infrastructures as discussed in the literature study section. So, we had to depend on the literature as well as SFPE values for this parameter. A seminal work on disabled people by this study \citep{Boyce1999} found that speed of assisted wheelchair on horizontal surface is 1.30 m/s, on ramps 0.89 m/s (upwards) and 0.96 m/s (downwards). Another experimental study \citep{Tsuchiya2007} on assisted wheelchair evacuation found that the speed on horizontal surface is 1.06 m/s. Based on these empirical studies and many different simulation studies as discussed in literature study section (section 2) in this paper, we have chosen the upper value of the disabled assisted evacuee speed as 1.19 m/s (default pathfinder value) and lower value as 0.89 m/s.
\subsubsection{Width of the staircase for underground metro layouts}
\ According to Delhi metro station layouts, the width of the stairs is 3 meters and the width of the escalator is 1.11 m. So, in our simulations for underground stations, we have taken the lower value of stair value as 4.11 meters. We need to see if an increase in the width of staircase could have effect on the total evacuation time of the occupants. So, we have taken the upper value of the stair width as 5.5 meters.
\subsubsection{Pre-evacuation time}
\ Pre-evacuation time is considered as the time before actual evacuation takes place after the fire is detected. This is the time when occupants decide whether to escape or not. They might also involve in any other activities. One study \citep{Shi2009} has listed different pre evacuation time for different types of buildings. However, according to different guidelines for metro station construction such as NFPA 130, the pre-evacuation time is considered to be 60 secs. In the literature review of metro station simulation (section 2), we found that almost all studies have taken pre-evacuation time as 60 secs. In some simulation studies, the pre-evacuation time is 90 secs or 120 secs for incorporating different classes of occupants (children, old) or different cognitive attitudes based on fire locations. Delhi metro follows NFPA 130 standards. So, in our study we have considered the lower value of pre-evacuation time as 60 secs and upper value as 90 secs.
\section{Design of Experiment and Simulation Results}
\ Design of experiments is a procedure to find out effects of many different factors on the response variable. This systemic procedure is used in Engineering (Process, Manufacturing, Chemical and Industrial) and Science to determine the effects of different factors on process yields. 
We have used the same concept in our simulation study. We found inconsistencies in literature in conducting evacuation simulations. In general, some questions appear which are not addressed in the literature.
What are the major factors that significantly affects the evacuation time?
How many simulations runs need to be conducted in order to get the significant parameters that affects the evacuation time?
In what order should the simulation runs be conducted to find out the significant parameters?
Whether 2nd and 3rd order interactions (between parameters) have any effect on the evacuation time?
To our knowledge no other studies have addressed these issues. We have used realistic parameters and design of experiments to address this issue. 
This section illustrates the design of experiments method used for different cases of simulation. There are many different types of designs for conducting experiments (full factorial, fractional factorial, Plackett-Burman etc.). These are discussed in details in the book by Montgomery \citep{Montgomery2006}. In our study we have used full factorial and different fractional factorial designs (one half and one quarter). When the number of factors is more, conducting a full factorial experiment may not be wise as the simulation runs will increase. Fractional factorial designs can relatively reduce the number of simulations runs and provide information about the significant parameters after they are analysed. Now, the different factorial designs and the results are discussed below for different cases. 
\subsection{Underground station Platform evacuation simulation (non-assisted)}
\ This subsection describes the design of experiment chosen for platform fire evacuation. There is no disabled occupant considered for this case.  The fire location is in the middle of the platform (near the middle stair). So, occupants use stairs and escalators at two ends of the platform (Refer Figure 1). The design used is a  fractional factorial design with resolution IV with no replications. Therefore, total number of simulations runs in the fractional factorial design=  =8.   Resolution IV designs are used to get the main effects un-confounded by two-factor interactions. Here the design generator used is D=ABC. We are using one half of the fractional factorial design. The four factors with two different levels are given in table 4. Negative sign (-1) indicates lower value and positive sign (+1) indicates higher value of the parameter. The design matrix and the results (evacuation times) of the underground non-assisted platform fire evacuation are shown in Table 5. Note that, NFPA critical evacuation time =240 secs, Delhi Metro critical evacuation time=330 secs
%%table4
\begin{table}
        \caption{Factors for non-assisted underground platform evacuation simulation}
        \centering
        \small
        {\begin{tabular}{|c|c|c|}
            \hline
                Factors & Low value (-1) & High value (+1)\\
            \hline
                Number of evacuee (A) & 350 & 800 \\
            \hline
                Speed of evacuee (B) & 1-1.5 m/s (uniform) & 1.5-2.15 m/s (uniform) \\
            \hline
                Stair width (C) & 3+1.11 m & 5.5 m \\
            \hline
               Pre-evacuation time (D) & 60 sec & 90 sec \\
            \hline
        \end{tabular}}
    \end{table}

%%table5
\begin{table}
        \caption{Design matrix and evacuation times for Underground non-assisted platform fire evacuation}
        \centering
        \small
        \begin{tabular}{|c|c|c|c|c|c|}
            \hline
                Std. Order & A & B & C & D=ABC & Evacuation Time (secs) \\
            \hline
                1 & -1 & -1 & -1 & -1 & 207.8 \\
            \hline
                2 & 1 & -1 & -1 & 1 & 259.0 \\
            \hline
                3 & -1 & 1 & -1 & 1 & 166.8 \\
            \hline
                4 & 1 & 1 & -1 & -1 & 198.8 \\
            \hline
                5 & -1 & -1 & 1 & 1 & 197.3 \\
            \hline
                6 & 1 & -1 & 1 & -1 & 232.3 \\
            \hline
                7 & -1 & 1 & 1 & -1 & 150.3 \\
            \hline
                8 & 1 & 1 & 1 & 1 & 183.3 \\
            \hline
        \end{tabular}
    \end{table}
So, it is observed from the simulation results that non-assisted platform evacuation for underground Delhi metro stations can be within the range of critical evacuation time value. Now, in order to estimate which of the factors (from Table 4) affect the evacuation time significantly, we have conducted ANOVA (analysis of variance) test. We have conducted no replication for each simulation run in order to keep the number of simulations less for computational purpose. The ANOVA results are inconclusive as we get no degrees of freedom for the error term. So, we have used a method of calculating the percentage contribution of each factor in the total sum of squares (SST). This method is discussed in the Montgomery book \citep{Montgomery2006}. The parameters which have less contribution will be dropped and we can get one degree of freedom for each parameter drop. Once the degrees of freedom for the error term is obtained, we can calculate the F-value and remark about the significance of each parameter. Table 6 shows ANOVA results and Table 7 shows the percentage contribution calculation for each parameter for this particular evacuation simulation case. From Table 7 it is evident that factor D (pre-evacuation time) is having less contribution and the interaction terms are also having less contribution. However, AB and AC interaction terms are not dropped as A, B and C are having more contribution. AD interaction term is dropped as D is dropped. Table 8 shows the modified ANOVA results for this underground non-assisted platform evacuation simulation. It is found from the results that number of evacuee (A), speed of evacuee (B) and Width of stair are significantly affecting the evacuation time for non-assisted underground platform evacuation of Delhi Metro. None of the interaction effects are significant in this case.

%%table6
\begin{table}
        \caption{The ANOVA results for underground non-assisted platform evacuation simulation}
        \centering
        \small
        \begin{tabular}{|c|c|c|c|}
            \hline
                Source & DOF & Adj. SS & Adj. MS  \\
            \hline
                Model & 7 & 8442.66 & 1206.09  \\
            \hline
                Linear & 4 & 8354.22 & 2088.56  \\
            \hline
                A & 1 & 2857.68 & 2857.68  \\
            \hline
                B & 1 & 4860.98 & 4860.98  \\
            \hline
                C & 1 & 598.58 & 598.58  \\
            \hline
                D & 1 & 36.98 & 36.98 \\
            \hline
                2-Way Interactions & 3 & 88.44 & 29.48 \\
            \hline
                AB & 1 & 56.18 & 56.18 \\
            \hline
                AC & 1 & 28.88 & 28.88 \\
            \hline
                AD & 1 & 3.38 & 3.38 \\
            \hline
                Error & 0 & - & - \\
            \hline
                Total & 7 & 8442.66 & - \\
            \hline
        \end{tabular}
    \end{table}
%%table7
\begin{table}
        \caption{Percentage contribution of each parameter to SST for underground non-assisted platform evacuation}
        \centering
        \small
        \begin{tabular}{|c|c|c|c|}
            \hline
                Factors & Adj. SS & percent contribution & Decision  \\
            \hline
                A & 2857.68 & 33.85 &  \\
            \hline
                B & 4860.98 & 57.58 &   \\
            \hline
                C & 598.58 & 7.09 &   \\
            \hline
                D & 36.98 & 0.44 & Dropped \\
            \hline
                AB & 56.18 & 0.67 & \\
            \hline
                AC & 28.88 & 0.34 &  \\
            \hline
                AD & 3.38 & 0.04 & Dropped \\
            \hline
                Total & 8442.66 &  &  \\
            \hline
        \end{tabular}
    \end{table}
    
\subsection{Underground station Platform evacuation simulation (assisted)}
\ This subsection discusses the assisted platform evacuation simulation for the underground station. We have five factors here as shown in Table 9. So, in order to reduce simulation runs, we have used  a fractional factorial design with resolution V and with no replications. Therefore, a total number of simulations runs in the fractional factorial design = 16. In resolution V designs the main effects un-confounded by three-factor interactions are estimated. The design generator is E=ABCD. We are using one-half of the fractional factorial design.  The results (evacuation times) of the underground assisted platform fire evacuation are shown in Table 10. It is evident from the results that for each case of the assisted underground platform evacuation simulation, the total evacuation time exceeds the critical value given by NFPA 130 (240 secs) and Delhi Metro (330 secs). From the effects chart (Pareto chart) analysis only factor B i.e., the speed of non-assisted evacuees is significantly affecting the total evacuation time. Figure 4 shows the Pareto chart for this simulation case. However, ANOVA results are inconclusive in this case like the earlier one due to the lack of degrees of freedom for the error term. The ANOVA results are shown in Table 11. Then we use the same percentage contribution method to get degrees of freedom for the error (factors less than 10 percent contribution is dropped) and finally remark about the significant parameters (main effects and interaction effects). Table 12 shows the percentage contribution method. Table 13 shows the final ANOVA results. The final results show that the Speed of non-assisted evacuees (B), speed of disabled evacuees (D), and a number of non-assisted (normal) and disabled evacuees together (interaction of AC) significantly influence the total evacuation time.
\begin{figure}
\centering
\includegraphics{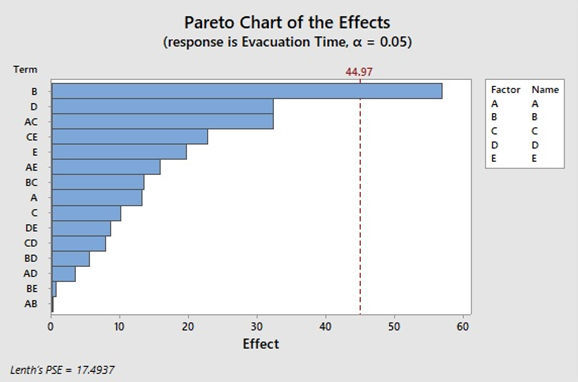}
\caption{Pareto chart for underground assisted platform evacuation simulation parameters}
\end{figure}
%%table8
\begin{table}
        \caption{Final ANOVA results for underground non-assisted platform evacuation}
        \centering
        \small
        \begin{tabular}{|c|c|c|c|c|c|c|}
            \hline
                Factors & Adj.SS & DOF & Adj.MS & F-value & F-table value & Decision \\
            \hline
                A & 2857.68 & 1 & 2857.68 & 141.60 & \multirow{7}{*}{F(1,2,0.05)= 18.51} & Significant\\ 
            \cline{1-7}
            B & 4860.98 & 1 & 4860.98 & 240.88 & & Significant\\
            C & 598.58 & 1 & 598.58 & 29.66 & & Significant\\
            AB & 56.18 & 1 & 56.18 & 2.78 & & Insignificant\\
            AC & 28.88 & 1 & 28.88 & 1.43 & & Insignificant\\
            Error & 40.36 & 2 & 20.18 & & & \\
            Total & 8442.66 & 7 & & & & \\
            \hline
        \end{tabular}
    \end{table}
%%table9
\begin{table}
        \caption{Factors for assisted underground platform simulation}
        \centering
        \small
        {\begin{tabular}{|c|c|c|}
            
            \hline
                Factors & Low value (-1) & High value (-1) \\
            \hline
                Number of evacuee (A) & 350 & 800\\
            \hline
                Speed of evacuee (B) & 1-1.5 m/s (uniform) & 1.5-2.15 m/s (uniform) \\
            \hline
               Number of disabled (C) & 4 & 8 \\
            \hline
              Speed of disabled (D) & 0.89 m/s  & 1.19 m/s \\
            \hline
              Pre-evacuation time (E) & 60 sec  & 90 sec \\
            \hline
        \end{tabular}}
    \end{table}
%%table10
\begin{table}
        \caption{Design matrix and evacuation times for Underground assisted platform evacuation}
        \centering
        \small
        \begin{tabular}{|c|c|c|c|c|c|c|}
            \hline
                Std. Order & A & B & C & D & E=ABCD & Evacuation Time (secs) \\
            \hline
                1 & -1 & -1 & -1 & -1 & 1 & 624.0 \\
            \hline
                2 & 1 & -1 & -1 & -1 & -1 & 551.5 \\
            \hline
                3 & -1 & 1 & -1 & -1 & -1 & 513.3 \\
            \hline
                4 & 1 & 1 & -1 & -1 & 1 & 542.3 \\
            \hline
                5 & -1 & -1 & 1 & -1 & -1 & 568.8 \\
            \hline
                6 & 1 & -1 & 1 & -1 & 1 & 616.0 \\
            \hline
                7 & -1 & 1 & 1 & -1 & 1 & 509.8 \\
            \hline
                8 & 1 & 1 & 1 & -1 & -1 & 545.5 \\
            \hline
                9 & -1 & -1 & -1 & 1 & -1 & 549.0 \\
             \hline
                10 & 1 & -1 & -1 & 1 & 1 & 566.8 \\
            \hline
                11 & -1 & 1 & -1 & 1 & 1 & 502.5 \\
            \hline
                12 & 1 & 1 & -1 & 1 & -1 & 452.0 \\
            \hline
                13 & -1 & -1 & 1 & 1 & 1 & 515.8 \\
            \hline
                14 & 1 & -1 & 1 & 1 & -1 & 577.3 \\
            \hline
                15 & -1 & 1 & 1 & 1 & -1 & 505.5 \\
            \hline
                16 & 1 & 1 & 1 & 1 & 1 & 543.3 \\
            \hline
        \end{tabular}
    \end{table}
\subsection{Underground station Train evacuation simulation (non-assisted)}
\ This subsection illustrates the underground train evacuation simulation. The fire is on the top of the train (middle section). This case is for non-assisted (normal) occupant evacuation simulation. For a train evacuation simulation, the number of evacuees is the occupants in the train plus the number of occupants on the platform. We have four factors as shown in Table 14. We have used  a fractional factorial design with resolution IV and with no replications.  
%%table11
\begin{table}
        \caption{The ANOVA results for underground assisted platform evacuation simulation}
        \centering
        \small
        \begin{tabular}{|c|c|c|c|}
            \hline
                Source & DOF & Adj. SS & Adj. MS  \\
            \hline
                Model & 15 & 28464.0 & 1897.6  \\
            \hline
                Linear & 5 & 19792.3 & 3958.5  \\
            \hline
                A & 1 & 702.2 & 702.2  \\
            \hline
                B & 1 & 12939.1 & 12939.1  \\
            \hline
                C & 1 & 406.0 & 406.0  \\
            \hline
                D & 1 & 4192.6 & 4192.6 \\
            \hline
                E & 1 & 1552.4 & 1552.4 \\
            \hline
                2-Way Interactions & 10 & 8671.7 & 867.2 \\
            \hline
                AB & 1 & 0.3 & 0.3 \\
            \hline
                AC & 1 & 4173.2 & 4173.2 \\
            \hline
                AD & 1 & 46.2 & 46.2 \\
            \hline
                AE & 1 & 1001.7 & 1001.7 \\
            \hline
                BC & 1 & 720.9 & 720.9 \\
            \hline
                BD & 1 & 119.9 & 119.9 \\
             \hline
                BE & 1 & 2.0 & 2.0 \\
             \hline
                CD & 1 & 244.9 & 244.9 \\
            \hline
                CE & 1 & 2070.2 & 2070.2 \\
            \hline
                DE & 1 & 292.4 & 292.4 \\
            \hline
                Error & 0 & - & - \\
            \hline
                Total & 15 & 28464.0 & - \\
            \hline
        \end{tabular}
    \end{table}
%%table12
\begin{table}
        \caption{Percentage contribution of each parameter to SST for underground assisted platform evacuation}
        \centering
        \small
        \begin{tabular}{|c|c|c|c|}
            \hline
                Factors & Adj. SS & percent contribution & Decision  \\
            \hline
                A & 702.2 & 2.47 & Dropped \\
            \hline
                B & 12939.1 & 45.46 &   \\
            \hline
                C & 406.0 & 1.43 & Dropped \\
            \hline
                D & 4192.6 & 14.73 & \\
            \hline
                E & 1552.4 & 5.45 & Dropped \\
            \hline
                AB & 0.3 & 0.0011 & Dropped\\
            \hline
                AC & 4173.2 & 14.66 &  \\
            \hline
                AD & 46.2 & 0.162 & Dropped \\
            \hline
                AE & 1001.7 & 3.52 & Dropped \\
            \hline
                BC & 720.9 & 2.53 & Dropped \\
            \hline
                BD & 119.9 & 0.421 & \\
            \hline
                BE & 2.0 & 0.007 & Dropped \\
            \hline
                CD & 244.9 & 0.86 & Dropped \\
            \hline
                CE & 2070.2 & 7.27 & Dropped \\
            \hline
                DE & 292.4 & 1.03 & Dropped \\
            \hline
                Total & 28464.0 &  &  \\
            \hline
        \end{tabular}
    \end{table}
The design generator is D=ABC and the total number of simulations runs are  =8. The design matrix and the simulation results are shown in Table 15. It is found from the results that most of the simulation results, in this case, exceed NFPA critical value (240 secs) and two of them exceed the Delhi Metro critical value (330 secs). The effects results are shown in the Pareto chart in Figure 5. From this figure, it is observed that only evacuee speed (factor B) is significant. Then we conducted ANOVA but the results are inconclusive due to a lack of degrees of freedom. So, we have used the percentage contribution method like the previous two cases and found significant parameters. The ANOVA results are shown in Table 16, percentage contribution calculations in Table 17 and final ANOVA results in Table 18. Factors having less than 5 percent contribution are dropped. Here, from table 17, AB and AC are not dropped because B and C are having more contributions and it is necessary to check whether these two factors in combination with any other factor have any significance on the total evacuation time.
%%table13
\begin{table}
        \caption{Final ANOVA results for underground assisted platform evacuation}
        \centering
        \small
        \begin{tabular}{|c|c|c|c|c|c|c|}
            \hline
                Factors & Adj.SS & DOF & Adj.MS & F-value & F-table value & Decision \\
            \hline
                B & 12939.1 & 1 & 12939.1 & 20.22 & \multirow{7}{*}{F(1,11,0.05) = 4.84} & Significant\\ 
            \cline{1-7}
            D & 4192.6 & 1 & 4192.6 & 6.55 & & Significant\\
            AC & 4173.2 & 1 & 4173.2 & 6.52 & & Significant\\
            BD & 119.9 & 1 & 119.9 & 0.19 & & Insignificant\\
            Error & 7039.2 & 11 & 639.93 & & & \\
            Total & 28464.0 & 15 & & & & \\
            \hline
        \end{tabular}
    \end{table}
After the final results, we are getting more parameters (parameter C) to be significant. So for Underground non-assisted train evacuation speed of evacuees (B) and width of stairs (C) are significantly affecting the total evacuation time. No other interaction terms are found to be significant.
%%table14
\begin{table}
        \caption{Factors for non-assisted underground train evacuation simulation}
        \centering
        \small
        {\begin{tabular}{|c|c|c|}
            
            \hline
                Factors & Low value (-1) & High value (-1) \\
            \hline
                Number of evacuee (A) & 1400+350 & 1650+350\\
            \hline
                Speed of evacuee (B) & 1-1.5 m/s (uniform) & 1.5-2.15 m/s (uniform) \\
            \hline
               Stair width (C) & 3+1.11 m & 5.5 m \\
            \hline
              Pre-evacuation time (D) & 60 sec  & 90 sec \\
            \hline
        \end{tabular}}
    \end{table}
%%table15
\begin{table}
        \caption{Design matrix and evacuation times for Underground non-assisted train fire evacuation}
        \centering
        \small
        \begin{tabular}{|c|c|c|c|c|c|}
            \hline
                Std. Order & A & B & C & D=ABC & Evacuation Time (secs) \\
            \hline
                1 & -1 & -1 & -1 & -1 & 352.0 \\
            \hline
                2 & 1 & -1 & -1 & 1 & 346.5 \\
            \hline
                3 & -1 & 1 & -1 & 1 & 256.3 \\
            \hline
                4 & 1 & 1 & -1 & -1 & 259.8 \\
            \hline
                5 & -1 & -1 & 1 & 1 & 308.3 \\
            \hline
                6 & 1 & -1 & 1 & -1 & 310.0 \\
            \hline
                7 & -1 & 1 & 1 & -1 & 227.3 \\
            \hline
                8 & 1 & 1 & 1 & 1 & 248.8 \\
            \hline
        \end{tabular}
    \end{table}
%%table16
\begin{table}
        \caption{The ANOVA results for underground non-assisted train evacuation simulation}
        \centering
        \small
        \begin{tabular}{|c|c|c|c|}
            \hline
                Source & DOF & Adj. SS & Adj. MS  \\
            \hline
                Model & 7 & 15432.5 & 2204.6  \\
            \hline
                Linear & 4 & 15047.4 & 3761.9  \\
            \hline
                A & 1 & 56.2 & 56.2  \\
            \hline
                B & 1 & 13170.6 & 13170.6  \\
            \hline
                C & 1 & 1806.0 & 1806.0  \\
            \hline
                D & 1 & 14.6 & 14.6 \\
            \hline
                2-Way Interactions & 3 & 385.1 & 128.4 \\
            \hline
                AB & 1 & 103.7 & 103.7 \\
            \hline
                AC & 1 & 79.4 & 79.4 \\
            \hline
                AD & 1 & 202.0 & 202.0 \\
            \hline
                Error & 0 & - & - \\
            \hline
                Total & 7 & 15432.5 & - \\
            \hline
        \end{tabular}
    \end{table}
%%table17
\begin{table}
        \caption{Percentage contribution of each parameter to SST for underground non-assisted train evacuation}
        \centering
        \small
        \begin{tabular}{|c|c|c|c|}
            \hline
                Factors & Adj. SS & percent contribution & Decision  \\
            \hline
                A & 56.2 & 0.36 & Dropped  \\
            \hline
                B & 13170.6 & 85.34 &   \\
            \hline
                C & 1806.0 & 11.70 &   \\
            \hline
                D & 14.6 & 0.095 & Dropped \\
            \hline
                AB & 103.7 & 0.672 & \\
            \hline
                AC & 79.4 & 0.514 &  \\
            \hline
                AD & 202.0 & 1.31 & Dropped \\
            \hline
                Total & 15432.5 &  &  \\
            \hline
        \end{tabular}
    \end{table}
%%table18
\begin{table}
        \caption{Final ANOVA results for underground non-assisted train evacuation}
        \centering
        \small
        \begin{tabular}{|c|c|c|c|c|c|c|}
            \hline
                Factors & Adj.SS & DOF & Adj.MS & F-value & F-table value & Decision \\
            \hline
                B & 13170.6 & 1 & 13170.6 & 144.84 & \multirow{7}{*}{F(1,3,0.05) = 10.13} & Significant\\ 
            \cline{6-6}
            C & 1806.0 & 1 & 1806.0 & 19.86 & & Significant\\
            AB & 103.7 & 1 & 103.7 & 1.14 & & Insignificant\\
            AC & 79.4 & 1 & 79.4 & 0.87 & & Insignificant\\
            Error & 272.8 & 3 & 90.93 & & & \\
            Total & 15432.5 & 7 & & & & \\
            \hline
        \end{tabular}
    \end{table}
\begin{figure}
\centering
\includegraphics{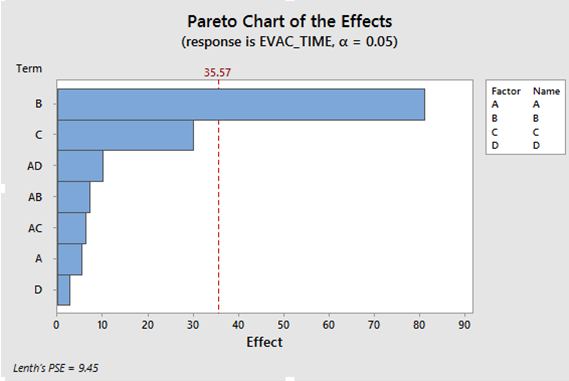}
\caption{Pareto chart for underground non-assisted train evacuation simulation parameters}
\end{figure}
\subsection{Underground station Train evacuation simulation (assisted)}
\ This subsection discusses the assisted platform evacuation simulation for underground station. We have six factors here as shown in Table 19. So, in order to reduce simulation runs we have used  fractional factorial design with resolution IV and with no replications. Therefore, total number of simulations runs in the fractional factorial design=  =16. The design generators are E=ABC and F=BCD. We are using one quarter fraction of the full factorial design. Using full factorial would require   runs. One quarter fraction reduces the number of simulations runs and saves the computational effort.  The design matrix and results (evacuation times) of the underground assisted train fire evacuation simulation are shown in Table 20. As we have so many results for so many different simulation cases, from this point onward we are not showing the inconclusive ANOVA results. We are only showing the percentage contribution and the final ANOVA results. Table 21 shows the percentage calculation and Table 22 shows the final ANOVA results for this simulation case. From the final ANOVA results all main effects except stair width is coming out to be significant. No interaction effects are found significant. For the elevated metro station simulations (which are discussed in the next section, stair width is not taken as a parameter for simulation as it is coming to be insignificant in most of the cases.
%%table19
\begin{table}
        \caption{Factors for assisted underground train simulation}
        \centering
        \small
        {\begin{tabular}{|c|c|c|}
            
            \hline
                Factors & Low value (-1) & High value (-1) \\
            \hline
                Number of evacuee (A) & 1400+350 & 1650+350\\
            \hline
                Speed of evacuee (B) & 1-1.5 m/s (uniform) & 1.5-2.15 m/s (uniform) \\
            \hline
               Number of disabled  (C) & 6 & 12 \\
            \hline
              Speed of disabled (D) & 0.89 m/s & 1.19 m/s \\
            \hline
              Stair width (E) & 3+1.11 m & 5.5 m \\
            \hline
              Pre-evacuation time (F) & 60 sec & 90 sec \\
            \hline
        \end{tabular}}
    \end{table}
%%table20
\begin{table}
        \caption{Design matrix and evacuation times for Underground assisted platform evacuation}
        \centering
        \small
        \begin{tabular}{|c|c|c|c|c|c|c|c|}
            \hline
                Std. Order & A & B & C & D & E=ABC & F=BCD & Evacuation Time (secs) \\
            \hline
                1 & -1 & -1 & -1 & -1 & -1 & -1 & 581.3 \\
            \hline
                2 & 1 & -1 & -1 & -1 & 1 & -1 & 596.8 \\
            \hline
                3 & -1 & 1 & -1 & -1 & 1 & 1 & 499.3 \\
            \hline
                4 & 1 & 1 & -1 & -1 & -1 & 1 & 769.8 \\
            \hline
                5 & -1 & -1 & 1 & -1 & 1 & 1 & 683.3 \\
            \hline
                6 & 1 & -1 & 1 & -1 & -1 & 1 & 845.3 \\
            \hline
                7 & -1 & 1 & 1 & -1 & -1 & -1 & 549.3 \\
            \hline
                8 & 1 & 1 & 1 & -1 & 1 & -1 & 545.3 \\
            \hline
                9 & -1 & -1 & -1 & 1 & -1 & 1 & 581.8 \\
            \hline
                10 & 1 & -1 & -1 & 1 & 1 & 1 & 572.8 \\
            \hline
                11 & -1 & 1 & -1 & 1 & 1 & -1 & 467.3 \\
            \hline
                12 & 1 & 1 & -1 & 1 & -1 & -1 & 437.3 \\
            \hline
                13 & -1 & -1 & 1 & 1 & 1 & -1 & 576.8 \\
            \hline
                14 & 1 & -1 & 1 & 1 & -1 & -1 & 664.0 \\
            \hline
                15 & -1 & 1 & 1 & 1 & -1 & 1 & 524.3 \\
            \hline
                16 & 1 & 1 & 1 & 1 & 1 & 1 & 602.8 \\
            \hline
        \end{tabular}
    \end{table}
%%table21
\begin{table}
        \caption{Percentage contribution of each parameter to SST for underground assisted train evacuation}
        \centering
        \small
        \begin{tabular}{|c|c|c|c|}
            \hline
                Factors & Adj. SS & percent contribution & Decision  \\
            \hline
                A & 20356 & 12.20 & \\
            \hline
                B & 31214 & 18.71 &   \\
            \hline
                C & 14683 & 8.80 & \\
            \hline
                D & 25865 & 15.50 & \\
            \hline
                E & 10440 & 6.26 & \\
            \hline
                F & 27332 & 16.38 & \\
            \hline
                AB & 220 & 0.13 & Dropped\\
            \hline
                AC & 368 & 0.22 & Dropped \\
            \hline
                AD & 6292 & 3.77 & Dropped \\
            \hline
                AE & 9443 & 5.66 & \\
            \hline
                AF & 11734 & 7.03 & \\
            \hline
                BD & 27 & 0.02 & Dropped \\
            \hline
                BF & 1101 & 0.66 & Dropped \\
            \hline
                ABD & 881 & 0.53 & Dropped \\
            \hline
                ABF & 6918 & 4.15 & Dropped \\
            \hline
                Total & 166874 &  &  \\
            \hline
        \end{tabular}
    \end{table}
%%table22
\begin{table}
        \caption{Final ANOVA results for underground assisted train evacuation}
        \centering
        \small
        \begin{tabular}{|c|c|c|c|c|c|c|}
            \hline
                Factors & Adj.SS & DOF & Adj.MS & F-value & F-table value & Decision \\
            \hline
                A & 20356 & 1 & 20356 & 9.01 & \multirow{7}{*}{F(1,7,0.05)= 5.59} & Significant\\ 
            \cline{6-6}
            B & 31214 & 1 & 31214 & 13.82 & & Significant\\
            C & 14683 & 1 & 14683 & 6.50 & & Significant\\
            D & 25865 & 1 & 25865 & 11.45 & & Significant\\
            E & 10440 & 1 & 10440 & 4.62 & & Insignificant\\
            F & 27332 & 1 & 27332 & 12.10 & & Significant\\
            AE & 9443 & 1 & 9443 & 4.18 & & Insignificant\\
            AF & 11734 & 1 & 11734 & 5.20 & & Insignificant\\
            Error & 15807 & 7 & 2258.14 & & & \\
            Total & 166874 & 15 & & & & \\
            \hline
        \end{tabular}
    \end{table}
\subsection{Elevated Delhi Metro station simulation}
\ Keeping in mind, the length of the paper, the design matrix, simulation results and the ANOVA test for parameter significance are not elaborately discussed for elevated Delhi Metro simulation cases. Table 23 discusses briefly the different cases (with different numbers of normal and disabled evacuees (with 2 different levels) as these are the main input parameters which change for different simulation cases while other parameter values remain same as in the previous simulation cases), their major outcomes and significant factors. Fire location for all cases are near to one of the two staircases of the platform level. So, typically one stair is blocked due to fire which is a critical case. 
It is observed from the results that only platform evacuation with no assisted evacuees do not cross the critical values given by NFPA and Delhi Metro planning standards. For assisted evacuation almost all the factors are significantly affecting total evacuation time. Pre-evacuation time is a major factor for elevated metro station evacuation.
%%table23
\begin{table}
        \caption{Different case results of elevated Delhi metro simulations}
        \centering
        \small
        \resizebox{\columnwidth}{!}{\begin{tabular}{|c|c|c|c|c|c|c|}
            \hline
                Simulation case & Number of evacuees & Number of disabled & Design used & Number of simulations runs & Critical value exceeds? & Significant parameters \\
            \hline
                Non-assisted train evacuation & 1000,1500 & 0 & Full factorial & 8 & Some cases & Evacuee number and speed of evacuee\\
            \hline
                Assisted train evacuation & 1000, 1500 & 8, 16 & Fractional factorial & 16 & All cases & Speed of normal evacuee, number of disabled and interaction between evacuee number and pre-evacuation time\\
            \hline
                Non-assisted platform evacuation & 250,500 & 0 & Full factorial & 8 & No cases & Evacuee Number, Speed of evacuee and pre-evacuation time\\
            \hline
                Assisted platform evacuation & 250, 300 & 3,6 & Fractional factorial & 16 & All cases & All parameters except number of normal evacuees\\
            \hline
        \end{tabular}}
    \end{table}
\section{Conclusions}
\ For any kind of experiment (laboratory, field or simulation) proper design of experiments are necessary.  Proper design can help estimate different main and interaction parameters. It can significantly reduce the computation expense for simulation studies.  Metro station evacuation studies are done in recent past. Most of the studies have inconsistencies in conducting simulation runs in standard order.  Significant effects are not properly estimated in any of the studies. Assisted evacuation is necessary for the disabled people who use metro as a mode of transportation. No studies in literature addressed this issue. There is no extensive literature on metro station simulations.  
This study has addressed all these issues. Firstly, this study has provided an extensive literature focusing on metro station simulations. It has also provided extensive literature on assisted evacuation simulation on other different infrastructures. This study has taken into account different fire scenarios (location of fire: different locations of platform) for two different layouts (underground and elevated) stations of Delhi metro. Basic fire locations are on the platform and on the train. So, we have train fire evacuation and platform fire evacuation. This study also has considered two basic simulation cases based on presence of disabled people in the simulation (assisted and non-assisted). So, we have eight different cases of simulations. We have used many different designs (full factorial and fractional factorial) for designing evacuation simulation experiment. 
 It is evident from the results that simple analysis (as done in most of the studies) fail to show many significant parameters. Design of experiments, ANOVA with percentage calculation can provide useful information about main factors and interaction (2-factor and 3-factor) factors that affect the total evacuation time. For underground non-assisted platform simulation all main factors except pre-evacuation time is significant. We have found different factors significant for different evacuation cases. For underground number of evacuees, number of disabled evacuee and both of their speeds are coming out to be more significant. Stair width is mostly significant for all cases of underground station simulations and pre-evacuation time is mostly insignificant for all cases except assisted train evacuation of underground station simulation. For elevated stations we found all factors significant for assisted evacuation cases except number of normal evacuees. 
From the results it is very much evident that presence of disabled people with need for assistance during evacuation significantly increases the total evacuation time. Efforts must be taken to provide infrastructure for quick mobilization of these groups. Efforts must be taken to increase the desired speed of the normal evacuees also. Special evacuation stairs with increased effective width can be provided in underground stations for evacuation purpose only. 
This study is the first one to use design of experiments for simulation purpose. It is quite evident that this method is very much useful for conducting simulation experiments. In future simulations can be done using design of experiments taking into account many other different behavioural parameters with varying levels to find out how those can affect the total evacuation time.

\bibliographystyle{unsrtnat}
\bibliography{Tarapada}%%Uncomment this line and comment out the ``thebibliography'' section below to use the external .bib file (using bibtex) 

%\begin{thebibliography}



% \end{thebibliography}

\end{document}